\documentclass[useAMS,usenatbib]{mn2e}
\usepackage{graphicx}
\usepackage{amssymb}
\usepackage{amsmath}
\usepackage{natbib}

\title[Microlensing by gas filaments]{Microlensing by gas filaments}
\author[V. Bozza and L. Mancini]{V. Bozza$^{1,2,3}$\thanks{E-mail:
valboz@sa.infn.it} and L. Mancini$^{2,3}$\thanks{lmancini@sa.infn.it}\\
$^1$Centro Studi e Ricerche ``Enrico Fermi'', via Panisperna 89/A,
Roma, Italy.\\
$^2$Dipartimento di Fisica ``E.R. Caianiello'', Universit\`{a} di
Salerno, I-84081 Baronissi (SA), Italy. \\
$^3$Istituto Nazionale di Fisica Nucleare, sez. Napoli, Italy.}

\begin{document}

\date{Received  / Accepted}

\maketitle

\label{firstpage}

\begin{abstract}
Gas in the interstellar matter is generally organized in
filamentary structures, which may be also relevant for a
complementary explanation of the dark matter in the Galactic halo.
We examine the possibility that such structures may act as
gravitational microlenses on background sources. To this purpose,
we derive the general properties of a cylindrical lens and compare
the light curves produced by such microlensing events with those
generated by spherically symmetric clouds. We find that the
establishment of the symmetry of the lens through the sole
analysis of the light curve may be problematic, while the analysis
of the astrometric shift of the centroid of the image can
discriminate between the two classes of clouds. On the basis of
our analysis, we find that only gas filaments with a very high
density could be detectable. Such clouds are unlikely to exist in
a long-lived state. Therefore, microlensing cannot discriminate on
the existence and the relevance of gas filaments in the Halo,
which could well be present and escape detection by ordinary
microlensing surveys.
\end{abstract}


\section{Introduction}

In the last years, the idea that baryonic dark matter in the halo
of our Galaxy could be in the form of self--gravitating cold gas
clouds has roused a considerable debate. This gas may arise in the
picture of a global fractal structure for the interstellar matter
in the Galactic halo
\citep{PCM,PC,DPIJR95a,DPIJR95b,DPIJR95c,DPIJR96} (see also
\citet{VSC} for a model of the ``fractalization''). The presence
of this gaseous component would help to account for the missing
dark matter in the halo of the Galaxy avoiding problems with the
initial mass function of compact objects helping to explain some
observational puzzles. However, it is difficult to establish the
existence and the relevance of such gas clouds on a clear
observational ground.

In order to check this hypothesis, gravitational microlensing by
gas clouds has been suggested by \citet{HenWid}. The lens masses
requested to give detectable events are similar to those of the
corresponding compact objects (MACHOs), i.e. of the order of a
tenth of the solar mass, and the size should be of the order of
the Einstein radius (few AU). They also suggest that these clouds
could be responsible for the extreme scattering events (ESE)
\citep{Fiedler}, which affect the radio emission of Quasars. This
hypothesis has been investigated and developed by \citet{WW98},
who pointed out that the ESEs could be explained by a Galactic
halo population of $\sim 10^{14}$ gas clouds, each of them having
an ionized envelope. In this way, the ESE are generated by the
motion across the line of sight of a lensing plasma which is able
in principle to amplify and deamplify an extragalactic point radio
source (see also \citet{W99,WWO}). \citet{DPIJR99a} have drawn
attention on the gamma ray emission of the gas clouds due to
cosmic ray scattering. This gamma ray could be recognized in
\citet{D98} detection according to \citet{DPIJR99b}. The details
of the gamma ray production from cosmic ray scattering have been
studied by \citet{Sciama}.

\citet{GS} have discussed the observational consequence of the
presence of cold gas clouds distributed in an extended, flattened,
three--dimensional halo. These authors have also considered
possible mechanisms to stabilize the gas clouds against
gravitational collapse and star formation (see also \citet{WW99}).
If the gas clouds were opaque, the microlensing surveys would
detect the occultation of the source stars \citep{GS,WW98}. On the
other hand, if the clouds were transparent at optical wavelengths,
classical refraction could supply an alternative amplification
mechanism \citep{Draine}. Moreover, a gaseous lens would impose
narrow infrared and far--red H$_{2}$ absorption lines on the
background stellar spectrum. Existing programs to observe
gravitational microlensing, supplemented by spectroscopy, can
therefore be used to either detect such events or constrain the
number of such gas clouds present in the Galaxy \citep{Draine}.
The magnification event rate has also been calculated for the case
of gaseous lensing of stars in the Large Magellanic Cloud (LMC) by
\citet{RD}.

\citet{BJMS} studied an intermediate possibility: if gas clouds
are effectively present in the halo, they could be likely
associated to massive compact halo objects (MACHOs). If this is
the case, these Pointlike Lenses Associated to Gas (PLAG) would be
missed by standard microlensing observations, because their light
curve would be significantly altered by the surrounding gas cloud.
On the other hand, isolated gas clouds could be quite difficult to
detect without a powerful amplification mechanism. In PLAGs,
gravitational lensing by the central object would provide this
mechanism, but secondary effects due to the cloud would show up in
a weak or strong fashion, depending on the density of the diffuse
matter. Significant differences in microlensing light curves of
background sources are present also in the case of binary systems
in a state of strong interaction, where mass exchanges, accretion
disks, common envelopes provide circumstellar matter which can
generate chromatic absorption effects \citep{BM}.

\citet{KBS} have examined the possibility that the data sets of
microlensing experiments searching for MACHOs can also be used to
search for occultation signatures by opaque gas clouds. \citet{DC}
examining the MACHO project light curves of 48 $\times 10^{6}$
stars towards the Galactic bulge, LMC and Small Magellanic Cloud
(SMC), have found no presence of dark cloud extinction events. On
an extragalactic ground, gas clouds could generate microlensing
signature on distant quasars \citep{TWH}. Up to now, no positive
detection has been reported by microlensing surveys on quasars
behind Virgo cluster.

So far, the relevance of gas clouds for gravitational microlensing
has been discussed only in the case of spherically symmetric
clouds \citep{HenWid}. However, in general, the gas dynamics tends
to form complex structures with lower symmetry, such as filaments
and sheets. These structures are typically induced by external
perturbations like shock waves from supernovae, which are the
prime engine for fragmentation and star formation in molecular
clouds \citep{Larson03}. The existence of a filamentary structure
within interstellar gas clouds has been supported by numerical
simulations and by many observations as well.

The theoretical studies of the structure of polytropic and
isothermal cylinders have been carried out by \citet{Ost1,Ost2}.
Stability analyses for the incompressible case have been performed
by \citet{ChaFer} and for the isothermal case by \citet{Stodol}.
The study of the fragmentation mechanism into filaments and then
into stars has been investigated by \citet{Larson85} and later on
by many authors. For a comprehensive review, see \citet{Larson03}.
The presence of magnetic fields and rotational motion play an
important role in the definition of privileged directions for the
formations of filaments, exerting always a slight stabilizing
effect against gravitational collapse \citep{Nak84,Nak95}. The
dynamical evolution of clouds is also influenced by the global
properties of the internal turbulence \citep{Klessen}. In any
case, the most efficient factor preventing a cloud from collapsing
is its internal temperature.

Filamentary structures appear to be present all around the Galaxy
at all scales and in all dynamical components, ranging from
molecular clouds with high rates of star formation
\citep{SchElm,Myers} to the regions surrounding the Galactic
center \citep{LaRosa}, from intermediate and high velocity clouds
in the Galactic halo \citep{Richter} to molecular structures in
the Magellanic Clouds \citep{Andre}. Even the various processes
that rule the gas circulation through the Milky Way halo, such as
the ``Galactic Fountains'' or matter debris torn from satellite
galaxies caused by Milky Way tidal forces (the Magellanic stream
is an example), rearrange the galactic gas preferably in
filamentary shapes.

From this general discussion, we draw two considerations. Firstly,
to investigate the observational effects of gaseous matter in the
Galaxy, it is mandatory to discuss clouds with cylindrical
symmetry, rather than spherical. Secondly, the actual limits of
the microlensing technique as a tool for the investigation of
diffuse interstellar matter still need to be established. At the
present time, it is not clear whether microlensing surveys should
expect events produced by gas clouds, and whether microlensing
statistics can be used for an investigation of the relevance of
diffuse components in the Galaxy. The absence of positive
detections is not significant, since a serious analysis should
start from a deep afterthought of the selection criteria.

To our knowledge, gravitational lensing by cylindrical lenses has
not been considered before, except for a study by \citet{BazDeF}.
The aim of the present work is to perform an analytical
investigation of this subject, with the purpose to establish the
minimal requirements to the parameters of a gas filament in order
to produce a detectable microlensing event. Besides this main
target, we will also compare gravitational lensing by gas
filaments with lensing by spherical clouds and cosmic strings.

The paper is structured as follows: in Sect. 2 we discuss the
general properties of the cylindrical lens. In Sect. 3 we compare
with the spherically symmetric case, showing that any microlensing
light curve generated by a spherically symmetric distribution can
be obtained by a cylindrical distribution, while the converse is
not true. In Sect. 4 we show that the degeneracy can be fully
broken using astrometric observations. In Sect. 5 we compare gas
filaments with cosmic strings from the gravitational lensing point
of view. Finally, in Sect. 6 we draw the conclusions, discussing
the relevance of microlensing for the investigation of diffuse
matter in the Galactic halo.

%
\section{The cylindrical lens}

As it is well known in the classical gravitational lensing theory,
if the size of the lens is much smaller than its distance from the
source and the observer, only the projected density distribution
is relevant for gravitational lensing. We define the critical
surface mass density as
\begin{equation}
\Sigma_{cr}=\frac{c^2D_{OS}}{4\pi G D_{OL}D_{LS}}
\end{equation}
and use a normalized surface density
$\kappa(\mathbf{x})=\Sigma(\mathbf{x})/\Sigma_{cr}$ \citep{SEF}.
As usual, $D_{OL}$, $D_{LS}$ and $D_{OS}$ indicate the distance
between observer and lens, lens and source, observer and source,
respectively.

A cylindrical lens with infinite length is projected to a density
distribution which is independent of one direction of symmetry.
Taking this direction to be the $x_2$ axis of the lens plane, the
generic density distribution takes the form
\begin{equation}
\kappa(x_1,x_2)=\lambda \sigma(x_1),
\end{equation}
where $\sigma(x)$ is an even function normalized in such a way
that
\begin{equation}
\int\limits_{-\infty}^{+\infty} \sigma(x)dx=1.
\end{equation}

Then $\lambda$ becomes the linear density of the filament, which
is supposed to be constant, or negligibly variable on the scales
relevant for microlensing.

The lens equation in the $x_2$ direction is trivial, since by
symmetry no deviation is expected. Working in normalized
coordinates \citep{SEF}, the deflection in the $x_1$ direction is
\begin{eqnarray}
&\alpha(x_1)&
=\frac{1}{\pi}\int\limits_{\mathbb{R}^2}\lambda\sigma(x'_1)
\frac{x_1-x'_1}{(x_1-x'_1)^2+(x_2-x'_2)^2} dx'_1 dx'_2 =\nonumber
\\&& = 2 \int\limits_{0}^{x_1}\lambda\sigma(x'_1) dx'_1.
\label{DefAng}
\end{eqnarray}

The lens equation is then simply
\begin{equation}
y_1=x_1-\alpha(x_1) \label{LensEq}
\end{equation}
with its Jacobian being
\begin{equation}
J=\left(
\begin{array}{cc}
1-2\lambda\sigma & 0 \\ 0 & 1
\end{array}
\right).
\end{equation}
Critical curves occur whenever
\begin{equation}
\kappa=\lambda \sigma=1/2. \label{EqCrit}
\end{equation}

They are described by the equation $|x_1|=x_c$, with $2\lambda
\sigma(x_c)=1$. By this equation, the critical curves are given by
lines parallel to the filament on each side. The caustics
corresponding to each critical line follow the equation
$y_1=y_c=\pm (x_c-\alpha_1(x_c))$, which still describes two lines
parallel to the filament but in the source plane.

Thanks to the high symmetry of this class of lens models, the
general theory has taken very few steps. It is however instructive
to visualize the main properties of the cylindrical models on a
specific example. To this purpose, we shall take for $\sigma(x_1)$
a gaussian profile with width $\sigma_0$
\begin{equation}
\sigma(x_1)= \frac{1}{\sigma_0\sqrt{2\pi}} e^{-x_1^2/2\sigma_0^2}.
\end{equation}

The deflection angle of such a density distribution is
\begin{equation}
\alpha(x_1)=\lambda \; \mathrm{Erf} \left(\frac{x}{\sigma_0
\sqrt{2}}\right). \label{DefAngGau}
\end{equation}

Strictly speaking, this formula is valid for $|x_1| \ll l$, where
$l$ is the length of the filament. In the opposite limit, when
$|x_1|\gtrsim l$ the integration on $x_2$ in Eq. (\ref{DefAng})
should be performed carefully, so that at very large $x_1$ we
recover the Schwarzschild lens behaviour. However, in the context
of this paper, we stick to the case $x_1 \ll l$, where we expect
the cylindrical lens effects to become important. In this limit,
the deflection angle is correctly represented by Eq.
(\ref{DefAng}).

\begin{figure}
\resizebox{\hsize}{!}{\includegraphics{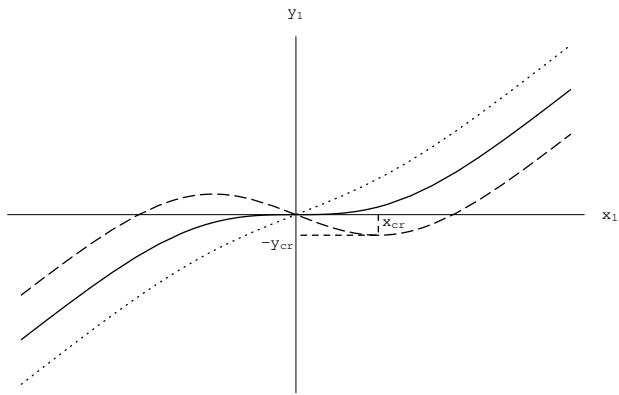}} \caption{The lens
equation for a cylindrical lens with a gaussian density profile
for different values of the linear density. The solid line has
$\lambda=\lambda_{cr}$, the dotted line $\lambda=0.5\lambda_{cr}$
and the dashed line $\lambda=1.5\lambda_{cr}$.}
 \label{Fig LensEq}
\end{figure}

In the particular case of a gaussian density profile, in Fig.
\ref{Fig LensEq} we plot the lens equation obtained using Eq.
(\ref{DefAngGau}) for different values of the linear density
$\lambda$. As the gaussian has its maximum at $x_1=0$, we can
define a critical linear density
\begin{equation}
\lambda_{cr}=\sigma_0 \sqrt{\frac{\pi}{2}},
\end{equation}
such that for $\lambda<\lambda_{cr}$ the Jacobian is always
positive and we have no critical curve, while for
$\lambda>\lambda_{cr}$ we have two critical curves. Since for a
cylindrical lens the Jacobian reduces to the derivative of the
lens equation, we can directly identify the critical curves with
the local extrema of the lens equation. For example, in Fig.
\ref{Fig LensEq}, the $\lambda=0.5 \lambda_{cr}$ curve is
monotonic, the $\lambda= \lambda_{cr}$ curve has a degenerate
stationary point in $x_1=0$ and the curve for $\lambda=1.5
\lambda_{cr}$ has two symmetric extrema at $x=\pm x_{cr}$, which
correspond to two critical curves on each side of the filament.
The height of the extrema determines the position of the
corresponding caustics $y_{cr}$. So, if $y$ is in the range
$|y|<y_{cr}$, we have three solutions for the lens equation, two
with positive parity at positions $|x|>x_{cr}$ and one in the
middle with negative parity. The parity can be directly read as
the sign of the derivative of the lens equation.

At this point we can draw some microlensing light curve samples
based on this filament model. What counts in a microlensing event
is the projected relative velocity between source and lens. As
usual, we set ourselves in the frame where the lens is fixed in
the sky and the source moves with respect to it. In our hypothesis
of uniform filament, we have an additional symmetry along the
$y_2$ axis, so that only the velocity component $v_1$ along $y_1$
is relevant. In practice, any microlensing event starts with a
source at $y_1=-\infty$, then develops with a source passing
through $y_1=0$ and will end up with the source at $y_1=+\infty$.

\begin{figure}
\resizebox{\hsize}{!}{\includegraphics{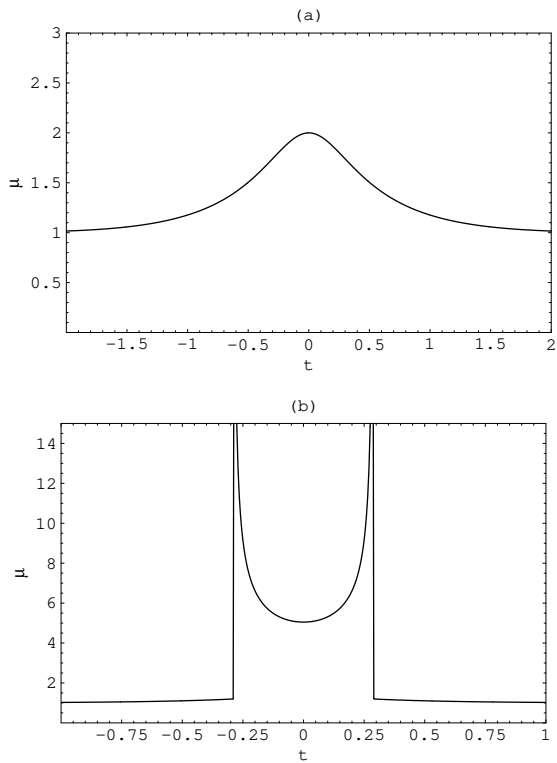}}
\caption{Microlensing light curves for a cylindrical lens with a
gaussian density profile with $\sigma_0=1$: (a) $\lambda=0.5
\lambda_{cr}$, (b) $\lambda=1.5 \lambda_{cr}$. Time is normalized
in such a way that $t=1$ corresponds to $y_1=1$.}
 \label{Fig LigCur}
\end{figure}

In order to draw a microlensing light curve, we have to solve the
lens equation (\ref{LensEq}) with $y_1=v_1 t$ and sum the
magnifications of all images, given by the absolute value of the
inverse of the Jacobian determinant evaluated at the image
positions. We can distinguish two cases: an underdense filament
$\lambda<\lambda_{cr}$ and an overdense filament
$\lambda>\lambda_{cr}$. In the first case (Fig. \ref{Fig LigCur}a)
we always have one image and the light curve is single--peaked
resembling the Paczynski curve or the light curves drawn by
\citet{HenWid}. In the second case (Fig. \ref{Fig LigCur}b) we
always have a caustic crossing event, as the caustic runs along
the whole filament. The light curve then is characterized by two
symmetric spikes and a central flat region, resembling the light
curves of binary lenses or those produced by diffuse clouds with
extended caustics.

\section{Comparison between the light curves generated by
cylindrical and spherical lenses} \label{Sect LightCurves}

Looking at the light curves shown in Fig. \ref{Fig LigCur} a very
important question arises: is it possible to identify the symmetry
of the lens (cylindrical or spherical) from any peculiarities of
the light curves? This question is not at all of academic
interest, but becomes of primary importance as soon as we want to
study the properties of the lensing object from the microlensing
light curves that we observe. The shape and the extension of the
clouds, together with their density are needed to estimate the
total mass of the object and, on a large statistic, the relevance
of the cloud population for the total matter in the halo.

In this section we will prove a simple theorem stating that any
light curve produced by a spherically symmetric lens can be
reproduced by a cylindrical lens with a suitable density profile.
The converse turns out to be not true. In fact, only a specific
class of cylindrical density profiles can be mimicked by
spherically symmetric ones.

Consider a spherical lens, with density profile $\sigma_s(x_s)$,
which is a function of the radial coordinate $x_s$ in the lens
plane. The spherical lens equation reads
\begin{equation}
y_s=x_s-\frac{m(x_s)}{x_s}, \label{LensEqS}
\end{equation}
where
\begin{equation}
m(x_s)=2\int\limits_0^{x_s}x'\sigma_s(x')dx'
\end{equation}
is the dimensionless mass enclosed in the radius $x_s$
\citep{SEF}.

The magnification of each image is given by
\begin{equation}
\mu_s(x_s)=\frac{1}{\frac{y_s(x_s)}{x_s}y'_s(x_s)}. \label{muS}
\end{equation}

Here we consider the case of a microlensing event with no caustic
crossing, hence with a single positive parity image. We refer the
reader to the appendix for the more complicated case of a caustic
crossing event.

We start by writing the source position as a function of time
\begin{equation}
y_s=\sqrt{b^2+v_s^2t^2},
\end{equation}
where $v_s$ is the relative velocity between source and lens and
$b$ is the impact parameter of the source trajectory with the
optical axis (the line connecting observer and lens).

The light curve is found inverting the lens equation
(\ref{LensEqS}) for each $y_s$ and evaluating the magnification on
the image position using (\ref{muS}).

On the other hand, the cylindrical lens is characterized by a
density distribution $\lambda\sigma_c(x_c)$, which is a function
of the coordinate $x_c$ transverse to the filament axis in the
lens plane. Its lens equation is
\begin{equation}
y_c=x_c-2\lambda\int\limits_0^{x_c} \sigma_c(x')dx' \label{LensC}
\end{equation}
and the magnification is
\begin{equation}
\mu_c(x_c)=\frac{1}{\frac{dy_c}{dx_c}}=\frac{1}{1-2\lambda
\sigma_c(x_c)}. \label{muC}
\end{equation}

In a microlensing event, as said before, the source position is
$y_c=v_c t$, where $v_c$ is the component of the source velocity
transverse to the filament axis.

Our aim is to show that for any microlensing event caused by a
spherically symmetric lens characterized by $\sigma_s$, $b$, $v_s$
there exist a microlensing event caused by a filament with some
$\sigma_c$ and $v_c$ such that the two light curves coincide.

The theorem can be stated by the equality between the
magnifications as functions of time
\begin{equation}
\mu_s(x_s(y_s(t)))=\mu_c(x_c(y_c(t))). \label{mumu}
\end{equation}

In both cases we have written the source position as function of
time. Eliminating $t$, we get a relation between the source
positions in the two microlensing events
\begin{equation}
y_s^2=b^2+r_v^2 y_c^2
\end{equation}
where $r_v=v_s/v_c$.

Solving with respect to $y_c$ and deriving with respect to $x_c$
we get
\begin{equation}
\frac{y_s}{r_v\sqrt{y_s^2-b^2}}\frac{dy_s}{dx_s}\frac{dx_s}{dx_c}=\pm
\frac{dy_c}{dx_c}.
\end{equation}
On the r.h.s. we have $1/\mu_c=1/\mu_s$. Substituting Eq.
(\ref{muS}) we finally get
\begin{equation}
\frac{dx_s}{dx_c}=\pm \frac{r_v}{x_s}\sqrt{y_s^2(x_s)-b^2},
\label{dxsdxc}
\end{equation}
where the plus sign holds for positive parity images and the minus
sign holds for negative parity ones. In a non--caustic crossing
event, only a positive parity image is present. We shall thus take
the plus sign.

This is a differential equation which gives us $x_s$ as a function
of $x_c$, relating the positions of the images in the two
microlensing events. If there were no lens, Eq. (\ref{dxsdxc})
would be solved by $x_s=\sqrt{b^2+r_v^2x_c^2}$. The presence of
the lens modifies this relation at low $x_c$, since when the
source has minimum impact parameter, its image is no longer just
$x_s=b$ but we have to solve the lens equation, finding
$x_s(0)=y_s^{-1}(b)$. This dictates the initial condition we need
to solve the differential equation (\ref{dxsdxc}). The r.h.s can
be explicitly expressed in terms of $x_s$ using the spherical lens
equation, so that for any density profile $\sigma_s$, any impact
parameter $b$ and velocity ratio $r_v$ we can get a family of
curves $x_s(x_c)$. These can be used to construct a filament
density profile using Eqs. (\ref{muC}) and (\ref{mumu})
\begin{equation}
\lambda \sigma_c(x_c)=\frac{1}{2}\left(1-\frac{1}{\mu_s(x_s(x_c))}
\right). \label{sigmac}
\end{equation}

\begin{figure}
\resizebox{10cm}{!}{\includegraphics{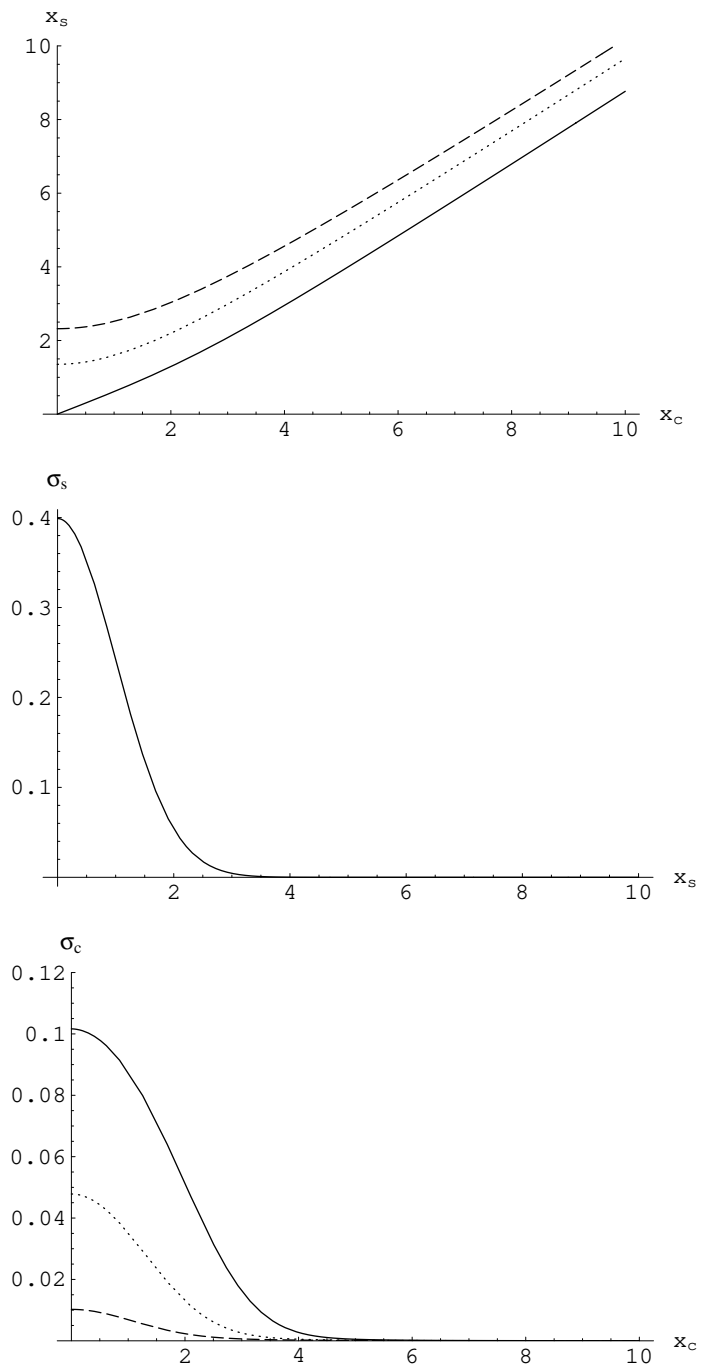}}
 \caption{Given a spherically symmetric gaussian density distribution
 shown in (b), (a) shows the relation between the spherical lens
 image position $x_s$ and the corresponding cylindrical lens image
 position $x_c$. (c) shows some cylindrical density profiles which
 reproduce the same microlensing event as (b) for different
 impact parameters: the solid line is for $b=0$, the dotted for
 $b=1$ and the dashed for $b=2$.
 }
 \label{Fig Eq1}
\end{figure}

In Fig. \ref{Fig Eq1} we show the results of this procedure
starting from a spherically symmetric lens with a gaussian density
profile (Fig. \ref{Fig Eq1}b). The equation which relates $x_s$ to
$x_c$ has been solved numerically and shown in Fig. \ref{Fig Eq1}a
for different impact parameters. Finally, the corresponding
density profiles for the filament are shown in Fig. \ref{Fig
Eq1}c. Of course, to a given spherically symmetric density profile
we cannot associate a unique equivalent cylindrical profile, but
rather a family of profiles which depend on the impact parameter
of the microlensing event we want to reproduce and the velocity
ratio of the two microlensing events. While the latter is only a
scale factor between $x_c$ and $x_s$, the impact parameter
determines the central density of the filament. The lower the
impact parameter, the higher the central density. For $b=0$, we
get the relation
\begin{equation}
\sigma_c(0)=\frac{1}{2\pi}\left[1-\left(1-\sigma_s(0) \right)^2
\right],
\end{equation}
which also tells us that for any subcritical spherically symmetric
density
\begin{equation}
\sigma_c(0)<\frac{\sigma_s(0)}{\pi}. \label{sigma0rel}
\end{equation}

So, looking at Fig. \ref{Fig Eq1}c we deduce  that any
microlensing event generated by a spherically symmetric
distribution can be explained by some filamentary distributions.
However, the latter need generically a lower central density to
generate the same peak magnification. The required central density
is suppressed even more as soon as $b$ departs from zero (see
discussion in Sect. \ref{Sect Discussion}).

A general property of spherically symmetric lenses is that the
tail of the event must follow a universal law
\begin{equation}
\mu_s\simeq 1+c_1 \, t^{-4}.
\end{equation}

This can be seen as follow. The Jacobian determinant of the
spherical lens is
\begin{equation}
\det J_s=\left(1-\frac{m(x_s)}{x_s^2} \right)
\left(1+\frac{m(x_s)}{x_s^2}-2\sigma_s(x_s) \right).
\end{equation}

Since the total mass of the distribution should be finite,
$m(x_s)$ should converge to a finite value $m(\infty)$. In order
for this to occur, $\sigma_s$ should drop faster than $x_s^{-2}$.
As a consequence, the dominant terms are
\begin{equation}
\det J_s \simeq 1- \frac{m^2(\infty)}{x_s^4}.
\end{equation}

At the tail of the event we have $x_s \simeq y_s \simeq v_s t$,
and then the magnification goes as
\begin{equation}
\mu_s \simeq 1+ \frac{m^2(\infty)}{v_s^4 t^4}.
\end{equation}

For the cylindrical lens we have no similar universal behaviour.
The tail of the event follows the tail of the distribution. From
Eq. (\ref{muC}), for large $t$ we find
\begin{equation}
\mu_c \simeq 1+2\lambda \sigma (v_c t).
\end{equation}

This suggests that only a  special class of cylindrical
microlensing events can have the correct form to be compatible
with a spherically symmetric explanation, namely the density
profiles falling as $x_c^{-4}$ at large distances from the
filament axis. For example, the light curves shown in Fig.
\ref{Fig LigCur} cannot be obtained by a spherically symmetric
lens. However, this general statement can be applied to physical
events only within the uncertainties, since a true spherically
symmetric event may have a first long tail coming from its diffuse
distribution which hides the last $t^{-4}$ tail behind the noise.

In the case of caustic crossing microlensing events, the search
for an equivalent cylindrical lens is complicated by the
creation/destruction of pairs of images. However, it can be indeed
proved that the theorem continues to hold. In the appendix, we
give an algorithm to generate a cylindrical density profile from a
caustic crossing microlensing event induced by a spherical cloud
which reproduces the same light curve.

\section{Astrometric microlensing by gas clouds}
\label{Sect Astrometry}

Astrometric measurements on the position of the centroid of the
star suffering microlensing were first proposed by \citet{Walker}
and \citet{HNP} (see also \citet{DomSah}). They are based on the
fact that even if we cannot resolve all the images individually,
if we have enough resolution (of the order of the Einstein radius
in the case of a pointlike lens), we can at least detect a shift
in the centroid of the lensed star, which follows the movement of
the images. In the case of a pointlike lens, it is well known that
the center of light (CoL) describes an ellipse in the sky, which
can be used for a better estimate of the microlensing parameters.

\begin{figure}
\resizebox{10cm}{!}{\includegraphics{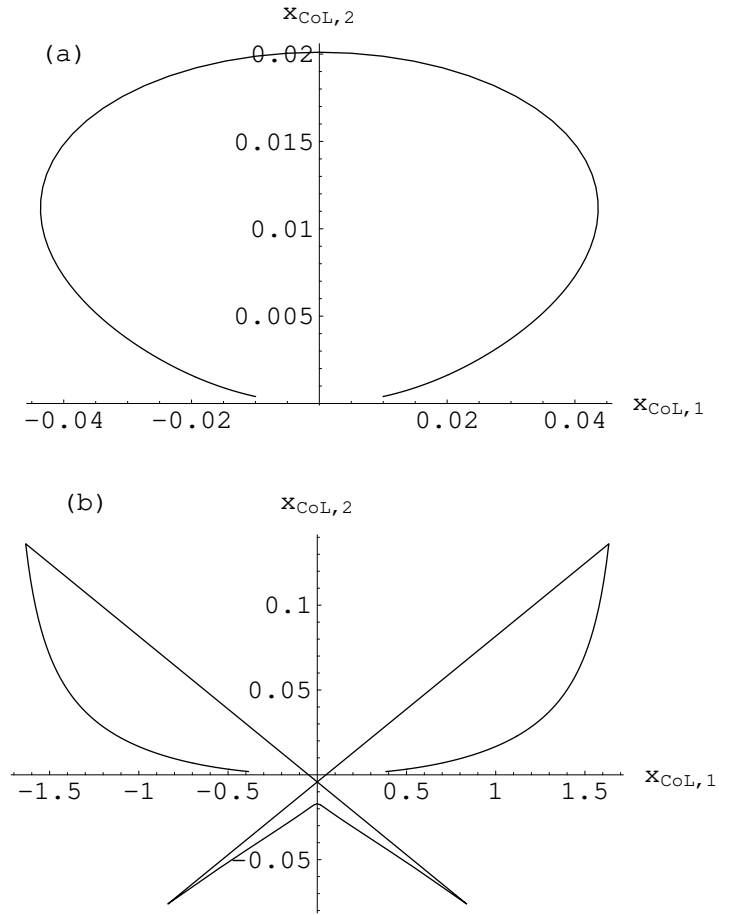}}
 \caption{Center of light trajectories for a spherical cloud with
 gaussian density. (a) A non--caustic crossing event. (b) A
 caustic crossing event.
 }
 \label{Fig CoLSph}
\end{figure}

Now let us consider a gas cloud as a lens with a spherical density
profile. The procedure to follow to calculate the CoL trajectory
is straightforward. We have to solve the lens equation for
$y=\sqrt{b^2+v_s^2 t^2}$, calculate the magnification $\mu^{(i)}$
of each image $x^{(i)}$ and evaluate a weighed average
\begin{equation}
x_{\mathrm{CoL}}=\sum \frac{x^{(i)}
\mu^{(i)}}{\mu_{\mathrm{tot}}}-y,
\end{equation}
where we have to subtract the position of the source in order to
get into the correct frame (the frame where the source is steady
and the lens moves). Up to now, the CoL curves for a diffuse
spherical lens have not been shown. In Fig. \ref{Fig CoLSph} we
show two examples obtained with a gaussian density profile. In
Fig. \ref{Fig CoLSph}a we show the CoL trajectory in a
microlensing event where the source does not cross any caustics.
In this case, we always have only one image and the CoL position
is just the difference between the image and the source position.
The trajectory is no longer an ellipse, but develops an asymmetry
between $t=0$ (the upper part of the curve) and $t=\pm \infty$
(the lower part). This asymmetry depends on the density profile
and can show up with an ellipse which is either flatter on the top
(like the one shown here) or flatter on the bottom.

In Fig. \ref{Fig CoLSph}b we show a case with a source which
enters a radial caustic. Two more images are generated on the side
opposite to the source, so that in the neighbourhood of $t=0$, the
CoL jumps in the lower half--plane (see \citet{GouHan} for similar
plots in binary microlensing).

\begin{figure}
\resizebox{10cm}{!}{\includegraphics{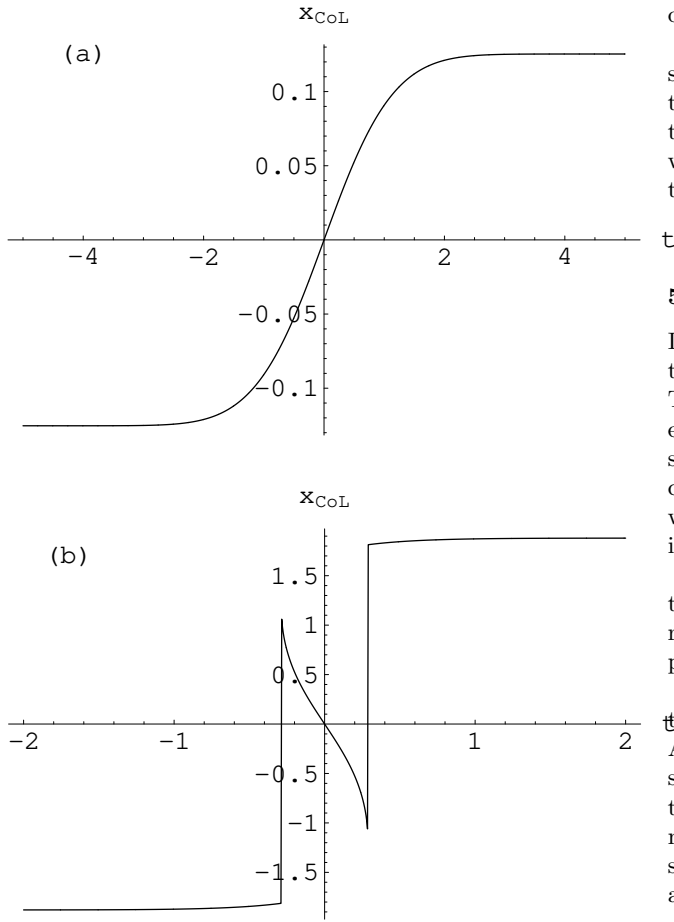}}
 \caption{Center of light trajectories for a filament with
 gaussian transverse density. (a) A non--caustic crossing event. (b) A
 caustic crossing event.
 }
 \label{Fig CoLCyl}
\end{figure}

Now let us consider a gas filament with a uniform linear density
and a gaussian transverse density profile. The big difference with
respect to the spherical lens is that now the CoL motion takes
place only along one direction, namely the direction transverse to
the filament. In Fig. \ref{Fig CoLCyl}a we show the motion of the
CoL as a function of time. Notice that, since we are considering
an infinite filament, the CoL does not return to zero at large
times. Actually, for a finite size filament this would obviously
occur, but at times very far from the luminosity peak. By the
cylindrical lens equation (\ref{LensEq}), we see that the starting
position for the CoL is $-\lambda$ and the final position is
$+\lambda$. So an observer looking at the star during the
microlensing event, would measure a shift equal to $2\lambda$ in
the direction transverse to the filament. This CoL motion is
absolutely different from the CoL motion of the spherical case.
Therefore the CoL motion provides a very clear criterion to
discriminate between the two kinds of symmetries.

In Fig. \ref{Fig CoLCyl}b we have a caustic crossing event. Like
in the spherically symmetric case, two more images are present in
the neighbourhood of $t=0$, which make the CoL jump on the
opposite side. The curve passes through 0 at $t=0$, where we have
two symmetric images on opposite sides of the filament and a third
image at the center.

%
\section{Gas Filaments vs. Cosmic Strings} \label{Sect Strings}
In the recent years, particular attention has been devoted to the
study of cosmic strings \citep{Zel,Vil}. These are topological
defects which may develop after high energy cosmological phase
transitions. To visualize a cosmic string, one may firstly
consider a circle on the plane. If we cut a sector from the circle
and glue the two cuts together, we obtain a conical surface. In a
similar way, a cosmic string is the cusp of a conical
three-dimensional space.

Since gaseous filaments have the same spatial symmetries as cosmic
strings, one may ask whether there are any relations or
similarities between the gravitational lensing properties of the
two classes of objects.

Following \citet{Got}, a cosmic string does not change the
curvature of the target spacetime, but only the topology. As a
consequence, light rays are not deflected by a cosmic string.
However, because of the conical structure of spacetime, two light
rays passing on different sides of the string may converge later
at the same arrival point. If a cosmic string lies between us and
a distant source, we see two images of the source separated by an
angle
\begin{equation}
\Delta \theta=8\pi \mu_s \sin \alpha \frac{D_{LS}}{D_{OS}},
\end{equation}
where $\mu_s$ is the string linear density and $\alpha$ is the
angle between the string and the direction from the observer to
the source. The two images are identical to the original source,
with no amplification or distortion. This is to be contrasted with
the gravitational lensing by gas filaments, where we have one or
three elongated images.

If we are in a typical microlensing situation, with a string
passing in front of the source and the two images unresolved, then
we can distinguish two cases, depending on the ratio between
$\Delta \theta$ and the source diameter. In fact, if this ratio is
larger than one, than the apparent luminosity of the source would
be rapidly doubled, as the second image appears. Then it would
stay constant for some time and finally return to the original
value, as the first image disappears.

If $\Delta \theta$ is smaller than the source size, then only a
portion of the source is doubled. The total amplification stays
smaller than two and we have a peculiar single peak curve, without
long tails, with a shape depending on the luminosity profile of
the source.

Summing up, we see that gravitational lensing by cosmic strings is
structurally different from classical gravitational lensing by
ordinary matter. It would be very difficult to bump into ambiguous
cases. For their peculiar properties, cosmic strings have been
invoked several times to explain twin objects showing no apparent
distortion \citep{TSB,SLC}.

%
\section{Discussion and Conclusions} \label{Sect Discussion}

Gas clouds have been invoked as a possible alternative explanation
for halo dark matter using ordinary baryonic constituents. Their
presence would represent a more orthodox solution to the problem
with respect to exotic particles and would solve some
observational puzzles. However, it is difficult to clearly prove
this hypothesis and to quantify its relevance in the dark matter
problem.

In this work we have revisited the original idea of microlensing
by spherical gas clouds by \citet{HenWid}, extending to the case
of gaseous filaments, which is the most likely geometry assumed in
a realistic picture. If we restrict ourselves to a light curve
analysis, then we proved in Sect. \ref{Sect LightCurves} that any
microlensing event generated by a spherically symmetric
distribution can be explained by a filamentary distribution, while
the converse is not true. The degeneracy between the two classes
of events can be broken by astrometric observations of CoL motion,
since this would be one-dimensional for cylindrical lenses and
two-dimensional for spherical ones.

A very important fact about filamentary clouds is that the study
of a microlensing event generated by them can be used for a
precise reconstruction of the density profile through Eqs.
(\ref{LensC}) and (\ref{muC}). This because any microlensing event
would probe the whole transverse structure of the filament,
including the center. In spherical clouds, instead, the presence
of an additional parameter (the impact parameter) does not allow a
univocal reconstruction of the density profile.

It is also important to remark that filaments are generally more
efficient microlenses than spherical clouds since, given a central
density of the cloud, a filament would give a higher magnification
for microlensing. In fact, the best magnification that can be
obtained by a spherical cloud is for a zero impact parameter
event. In this case, Eq. (\ref{sigma0rel}) shows that spherically
symmetric clouds require a slightly larger central density than
cylindrical ones in order to give the same magnification. Of
course, for larger impact parameters the difference would increase
even more.

Similarly to what happens for spherical clouds
\citep{HenWid,Draine,BJMS}, cold gas causes reddening of the light
of the background source due to absorption by Rayleigh scattering.
As a consequence, a high chromaticity in the light curve could be
expected. Spectroscopic measurements during these microlensing
events would yield a precise knowledge of the chemical composition
of the clouds. Of course, such detailed analysis could be
performed in a completely renewed framework for microlensing
surveys, with selection criteria completely redisegned to unveil
microlensing events generated by gas clouds.

As regards the microlensing probability in a given model of gas
distribution, we need to specify three parameters: the total mass
in gas clouds, the central density $\rho_0$ and the typical
transverse size $\sigma_0$ of a single cloud. This is to be
contrasted with the MACHO picture, where the last two parameters
are replaced by a single one, namely the mass of the MACHO. Of
course, in any realistic model of the baryonic halo, we should
replace the mass of the MACHO with a mass function. In the same
way, for an ensemble of gas clouds, we would have a distribution
$dN/d\rho_0 d\sigma_0$, rather than a single value for $\rho_0$
and $\sigma_0$. An important fact is that one can easily find that
the total microlensing probability is independent of the shape of
the cloud. So, for example, the frequency of microlensing events
generated by a population of filamentary clouds would be exactly
the same as for a population of spherical clouds with the same
total mass and similar transverse size. This allows us to neglect
higher order shape parameters to determine the microlensing
probability.

Now let us analyze the minimal requirements for the parameters of
a gas filament in order to be detectable by present microlensing
surveys. A typical microlensing campaign puts a threshold on the
minimal amplification and has a detection efficiency limited to
events with duration within a specified range. For a typical event
with maximum amplification at the peak $\mu_{max}=2$, we shall now
derive a relation between the duration and the central density.
For our Gaussian density profile, we can easily write down the
following relation between the central volume density $\rho_0$,
the linear density $\lambda$ and the radius of the filament
$\sigma_0$
\begin{equation}
\rho_0= \frac{\lambda}{2\pi \sigma_0^2}.
\end{equation}

By Eq. (\ref{muC}), we can immediately deduce the radius of the
filament as a function of the central density and magnification at
the peak
\begin{equation}
\sigma_0=\frac{1}{2\rho_0\sqrt{2\pi}}\left(1-\frac{1}{\mu_{max}}
\right).
\end{equation}

The semi-duration of the event can be quantified by $t_{1/2}$,
which is the time needed to halve the amplification. It is defined
by the relation
\begin{equation}
\mu(x_1(t_{1/2}))=\frac{\mu_{max}+1}{2}.
\end{equation}

By the lens equation (\ref{LensEq}),
\begin{equation}
t_{1/2}=\frac{1}{v} \left[x_{1/2}-\alpha (x_{1/2}) \right],
\end{equation}
where $v$ is the relative projected velocity between source and
filament and $x_{1/2}$ is defined by
\begin{equation}
\mu(x_{1/2})=\frac{\mu_{max}+1}{2}.
\end{equation}

Then, by simple algebra, using (\ref{DefAngGau}) and restoring
dimensional units we obtain
\begin{eqnarray}
& t_{1/2}=& \frac{k_0 \Sigma_{cr}}{v \rho_0} \\ %
& k_0=& \frac{\mu_{max}-1}{4\sqrt{\pi} \mu_{max}^2} \left[
\sqrt{\pi} (\mu_{max}-1) \mathrm{Erf} \left( \sqrt{\ln \frac{
\mu_{max}+1}{\mu_{max}}} \right) \right. \nonumber \\
&& \left. -2\mu_{max} \sqrt{\ln \frac{ \mu_{max}+1}{\mu_{max}}}
\right]=0.05.
\end{eqnarray}

Fixing the maximum semi-duration to 150 days, the projected
velocity to 200 km/s, and the chemical composition of the filament
to H$_2$ molecules, we can calculate the central volume density of
the filament for different observational targets (the geometry
enters through the critical density $\Sigma_{cr}$). For example,
if the source is a star in the LMC ($D_{OS}=50$ kpc), the number
density at the center of the filament should be at least
$n_0=1.2\times 10^{12}$ cm$^{-3}$. If the source is in M31
($D_{OS}=700$ kpc), we have $n_0=8.7\times 10^{10}$ cm$^{-3}$. If
the source is a quasar at 1 Gpc distance, we have $n_0=6\times
10^{7}$ cm$^{-3}$. Of course, leaving fixed the magnification at
the peak, which is a function of the product $\rho_0 \sigma_0$, we
can lower the central density increasing the radius of the
filament. However, the duration of the event is proportional to
the inverse of $\rho_0$ and would rapidly become too long for a
reasonable microlensing campaign. We can repeat the calculation
changing the maximum amplification $\mu_{max}$, but this only
changes the final result by a numerical factor of order unity.

According to \citet{HenWid} the typical molecular number density
for spherical clouds required to generate appreciable
gravitational lensing effects is even higher ($10^{15}$ cm$^{-3}$)
than what we have found. The higher densities experimentally
measured for dark clouds are of the order of $10^5$ cm$^{-3}$
\citep{CarOst}, while in massive star forming cloud, they may
exceed $10^7$ cm$^{-3}$ in some locations \citep{Evans,GarLiz}.
Only quasar microlensing seems to have enough sensitivity to probe
diffuse interstellar matter on the basis of present knowledge.
This potentially interesting topic would thus deserve a separate
study beyond the scope of this work.

As regards stellar microlensing on Galactic or Local-Group scales,
the situation is not really promising. In fact, it is difficult to
imagine a mechanism which keeps in a metastable state a gas cloud
with density of the order of $10^{10}$ cm$^{-3}$ for a
sufficiently long life. Such density can only be reached during a
collapsing stage for a relatively short time, even if global
properties like magnetic field and internal turbulence partially
contrast the gravitational collapse of the gas \citep{Klessen}.

We conclude that only quasar microlensing can hope to say
something on the presence and the relevance of diffuse clouds or
filaments in the haloes of other galaxies. Galactic microlensing
campaigns cannot give any definite answer to the question on the
existence of a relevant diffuse baryonic component in the Galactic
halo, unless completely new ideas intervene in the observational
strategies or high density gas clouds are effectively sustained by
some still unknown mechanisms. A positive detection of a
microlensing event due to a purely diffuse cloud or filament would
be a quite unrealistic chance, which could not help us to
understand the global problem all the same, since it would be
generated by a very peculiar object. On the other hand, the
absence of signs of diffuse matter in present microlensing
campaigns does not exclude the possibility of large distributions
of filaments in the Halo, which would be too diluted to give
appreciable microlensing effects. One possibility still remaining
open is for gas clouds associated to compact objects \citep{BJMS},
which would provide an efficient microlensing amplification
mechanism and signal the presence of gas in the Halo. Otherwise,
the problem should definitely be tackled by other techniques.

\section*{Acknowledgments}

We thank the CERN theory department and the Institute for
theoretical physics of the Z\"urich University for hospitality. We
also thank Gaetano Scarpetta for careful reading of the
manuscript.

\appendix
\section{Comparison between the light curves generated by
cylindrical and spherical lenses: Caustic crossing events}

In this appendix we extend the theorem of Sect. \ref{Sect
LightCurves} to caustic crossing microlensing events. Suppose we
have a spherically symmetric distribution which exceeds the
critical surface density in a neighbourhood of its center. Then we
shall have a tangential critical curve when $m(x_s)=x_s^2$ and an
inner radial critical curve when $dy_s/dx_s=0$. The caustic
corresponding to the tangential critical curve is just a point on
the optical axis, while the caustic corresponding to the radial
critical curve is a circle of finite size, which can be crossed by
a source with a sufficiently small impact parameter. Here we shall
consider the case of a single radial critical curve.

In the spherically symmetric case, we have a principal image which
is not involved in creation/destruction processes. Then, when the
source enters the caustic, an additional pair of image forms: one
in the range between the tangential critical curve and the radial
critical curve and another between the radial critical curve and
the origin.

In the cylindrical case, we have one image on the same side of the
source when this is outside the caustic. Then, when the source
enters the caustic, two images are generated on the opposite side:
one moves away from the filament and the other moves towards the
center of the filament, both starting from the critical line. When
the source is aligned with the filament, we have a symmetric
situation where the original image and the second one are outside
the critical lines on specular positions and the third one is
right at the center.

When we have more than one image, in principle we can distribute
the total magnification among the images in arbitrary proportions.
For each choice we can find a different cylindrical density
profile. Just to prove the existence of at least one profile, we
impose that to any image in the spherically symmetric case, there
corresponds one image in the cylindrical case which has the same
magnification. Our choice is to identify the most external images
each other and to identify the two remaining ones by their parity.
Even with this simple choice, the procedure to find $x_s(x_c)$ is
somewhat complicated.

Eq. (\ref{dxsdxc}) can still be used to relate the positions of
the images, though we should take care of the sign of the images.
To obtain the three images, we have to start by suitable initial
conditions. In fact, when the source reaches its minimum distance
$b$ from the center, the spherical lens equation yields three
solutions that we can label in decreasing order as $x_s^{(i)}$
with $i=1,2,3$. For the negative parity images the logical choice
is $x_s(0)=x_s^{(2)}$. The differential equation (\ref{dxsdxc})
with the minus sign can then be followed without singularities
even through the radial critical curve until $y_s$ becomes equal
to $b$. This happens when $x_s$ reaches $x_s^{(3)}$. This
practically means that we obtain a unique function $x_s(x_c)$
which covers not only the negative parity image, but also the
second positive parity one, passing through the critical point
$(x_c^{cr},x_s^{cr})$, which can be used to fix the position of
the critical line in the cylindrical lens. This branch dies in
$x_c^{dis}$ where $x_s$ reaches $x_s^{(3)}$. Finally, the
principal image can be found solving Eq. (\ref{dxsdxc}) with the
positive sign and the initial condition
$x_s(x_c^{dis})=x_s^{(1)}$. These choices correspond to the
identification explained above between the images of the two
frames. At the end, by Eq. (\ref{sigmac}), we get the final
density profile for the cylindrical lens as a function of $x_c$.

\begin{figure}
\resizebox{10cm}{!}{\includegraphics{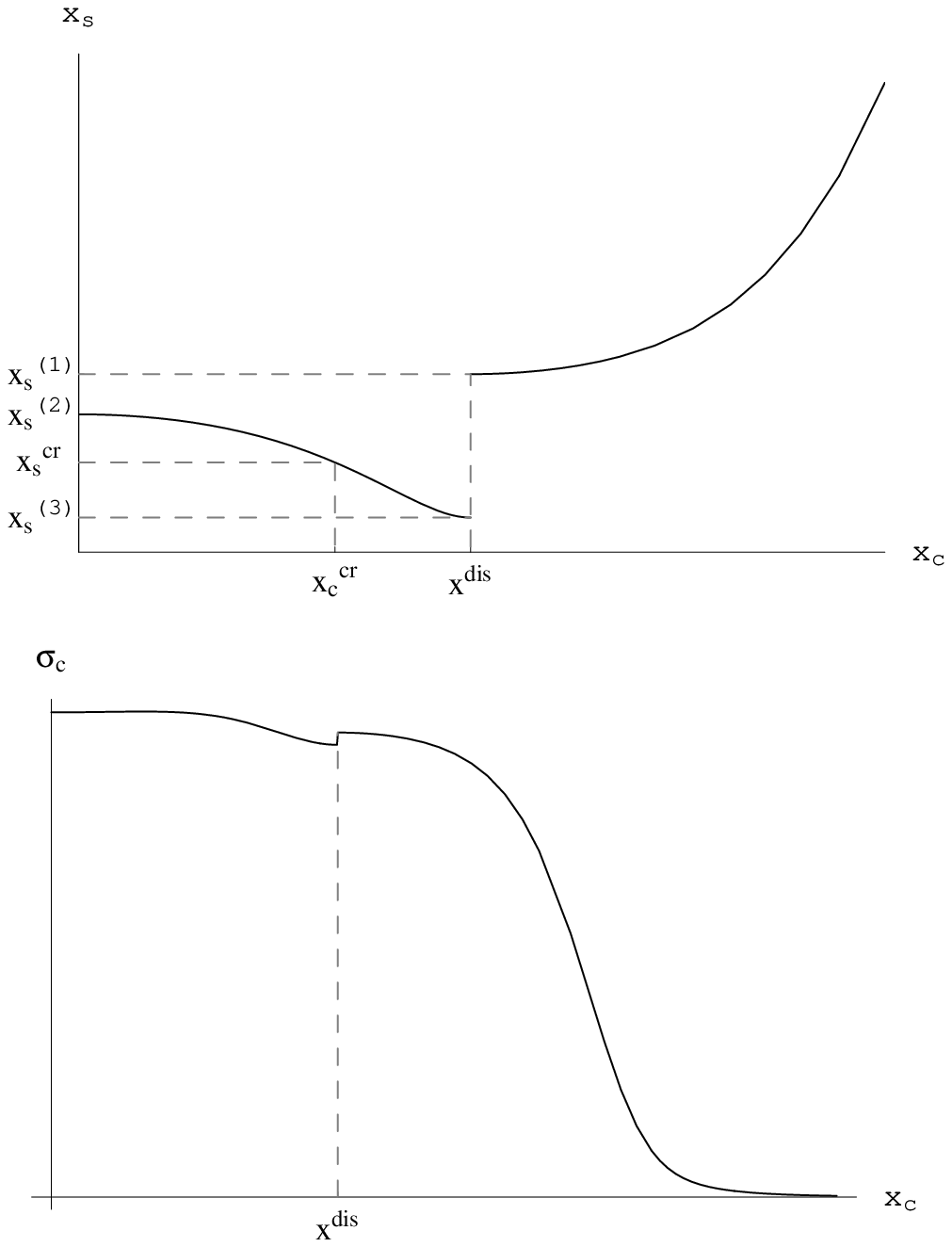}}
 \caption{
The same as figure \ref{Fig Eq1} but starting from a higher
spherical lens density and $b=0.05$, in order to have caustic
crossing and multiple image formation. (a) Relation between the
spherical lens image positions $x_s$ and the corresponding
cylindrical lens image positions $x_c$. (b) The obtained
cylindrical density profiles which reproduces the same caustic
crossing microlensing event as the original spherically symmetric
lens.
 }
 \label{Fig Eq2}
\end{figure}

In Fig. \ref{Fig Eq2}a we plot the relation between the positions
of the images. Here we see that our choice has generated a
relation which is necessarily discontinuous in $x_c=x_c^{dis}$.
Then the outermost image is mapped in the sector
$x_c>x_c^{dis}$,$x_s>x_s^{(1)}$, the negative parity image in
$0<x_c<x_c^{cr}$,$x_s^{cr}<x_s<x_s^{(2)}$ and the third image in
the sector $x_c^{cr}<x_c<x_c^{dis}$,$x_s^{(3)}<x_s<x_s^{cr}$. As a
consequence, also the cylindrical density profile is
discontinuous. A simple way to smooth the discontinuity would be
to allow the images of the cylindrical lens to carry a different
weight in the total magnification than their corresponding
spherical lens images.

\label{lastpage}


\begin{thebibliography}{}
%
\bibitem[Alcock et al.(2000)]{AAA}
Alcock C. et al., 2000, ApJ, 542, 281
%
\bibitem[Andr\'{e} et al.(2004)]{Andre}
Andr\'{e} M. K. et al., 2004, astro-ph/0404063
%
\bibitem[Bazin \& De Freitas(1987)]{BazDeF}
Bazin M. J., De Freitas L., 1987, Ap\&SS, 138, 381
%
\bibitem[Bozza et al.(2002)]{BJMS}
Bozza V., Jetzer Ph., Mancini L., Scarpetta, G., 2002, A\&A, 382,
6
%
\bibitem[Bozza \& Mancini(2002)]{BM}
Bozza V., \& Mancini L., 2002, A\&A, 394, L47
%
\bibitem[Carroll \& Ostlie(1996)]{CarOst} Carroll B.W., Ostlie D.A., 1996, ``An
introduction to modern astrophysics'' (Addison-Wesley Pub. Comp.)
%
\bibitem[Chandrasekhar \& Fermi(1953)]{ChaFer}
Chandrasekhar S., Fermi E., 1953, ApJ, 118, 116
%
\bibitem[De Paolis et al.(1995a)]{DPIJR95a}
De Paolis F., Ingrosso G., Jetzer Ph., Roncadelli M., 1995, PRL,
74, 14
%
\bibitem[De Paolis et al.(1995b)]{DPIJR95b}
De Paolis F., Ingrosso G., Jetzer Ph., Roncadelli M., 1995, A\&A,
295, 567
%
\bibitem[De Paolis et al.(1995c)]{DPIJR95c}
De Paolis F., Ingrosso G., Jetzer Ph., Qadir A., Roncadelli M.,
1995, A\&A, 295, 647
%
\bibitem[De Paolis et al.(1996)]{DPIJR96}
De Paolis F., Ingrosso G., Jetzer Ph., Roncadelli M., 1996,
Ap\&SS, 235, 329
%
\bibitem[De Paolis et al.(1999a)]{DPIJR99a}
De Paolis F., Ingrosso G., Jetzer Ph., Roncadelli M., 1999, ApJ,
510, L103
%
\bibitem[De Paolis et al.(1999b)]{DPIJR99b}
De Paolis F., Ingrosso G., Jetzer Ph., Roncadelli M., 1999,
astro-ph/9906083
%
\bibitem[de Vega, Sanchez \& Combes(1996)]{VSC}
de Vega H. J., S\'anchez N., Combes F., 1996, PRD, 54, 106008
%
\bibitem[Dixon et al.(1998)]{D98}
Dixon D. D. et al., 1998, Nat, 3, 539
%
\bibitem[Dominik \& Sahu(2000)]{DomSah}
Dominik M., Sahu K., 2000, ApJ, 534, 213
%
\bibitem[Draine(1998)]{Draine}
Draine B. T., 1998, ApJ, 509, L41
%
\bibitem[Drake \& Cook(2003)]{DC}
Drake A. J., Cook K. H., 2003, ApJ, 589, 281
%
\bibitem[Evans(1999)]{Evans}
Evans N. J., 1999, Ann. Rev. Astron. Astroph., 37, 311
%
\bibitem[Fiedler et al.(1994)]{Fiedler}
Fiedler R., Dennison B., Johnston K. J. et al., 1994, ApJ, 430,
581
%
\bibitem[Garay \& Lizano(1999)]{GarLiz}
Garay G., Lizano S., 1999, Astron. Soc. Pacific, 111, 1049
%
\bibitem[Gerhard \& Silk(1996)]{GS}
Gerhard O., Silk J., 1996, ApJ, 472, 34
%
\bibitem[Gott(1985)]{Got}
Gott J. R., 1985, ApJ, 288, 422
%
\bibitem[Gould \& Han(2000)]{GouHan}
Gould A., Han C., 2000, ApJ, 538, 653
%
\bibitem[Henriksen \& Widrow(1995)]{HenWid}
Henriksen R. N., Widrow L. M., 1995, ApJ, 441, 70
%
\bibitem[H{\o}g, Novikov \& Polnarev(1995)]{HNP}
H{\o}g E., Novikov I. D., Polnarev A. G., 1995, A\&A, 294, 287
%
\bibitem[Kerins, Binney \& Silk(2002)]{KBS}
Kerins E., Binney J., Silk J., 2002, MNRAS, 332, L29
%
\bibitem[Klessen \& Ballesteros--Paredes(2004)]{Klessen}
Klessen R. S., Ballesteros--Paredes J., 2004, astro-ph/0402038
%
\bibitem[Lasserre et al.(2000)]{LAA}
Lasserre T., Afonso C., Albert J.N. et al., 2000, A\&A, 355, L39
%
\bibitem[La Rosa  et al.(2004)]{LaRosa}
La Rosa T. N., Nord M. E., Lazio T. J., Kassim N. E., 2004,
astro-ph/0402061
%
\bibitem[Larson(1985)]{Larson85}
Larson R. B., 1985, MNRAS, 214, 379
%
\bibitem[Larson(2003)]{Larson03}
Larson R. B., 2003, Rep. Prog. Phys., 66, 1651
%
\bibitem[Myers et al.(1991)]{Myers}
Myers P. C., Fuller G. A., Goodman A. A., Benson P. J., 1991, ApJ,
376, 561
%
\bibitem[Nakamura(1984)]{Nak84}
Nakamura T., 1984, Prog. Theor. Phys., 70, 212
%
\bibitem[Nakamura et al.(1995)]{Nak95}
Nakamura T., Hanawa T., Nakano T., 1995, ApJ, 444, 770
%
\bibitem[Ostriker(1964)]{Ost1}
Ostriker J., 1964, ApJ, 140, 1056
%
\bibitem[Ostriker(1965)]{Ost2}
Ostriker J., 1965, ApJ Suppl., 11, 167
%
\bibitem[Pfenniger \& Combes(1994)]{PC}
Pfenniger D., Combes F., 1994, A\&A, 285, 94
%
\bibitem[Pfenniger, Combes \& Martinet(1994)]{PCM}
Pfenniger D., Combes F., Martinet L., 1994, A\&A, 285, 79
%
\bibitem[Rafikov \& Draine(2001)]{RD}
Rafikov R. R., Draine B. T., 2001, ApJ, 547, 207
%
\bibitem[Richter et al.(2003)]{Richter}
Richter P., Sembach K. R., Howk J. C., 2003, A\&A, 405, 1013
%
\bibitem[Sazhin et al.(2003)]{SLC}
Sazhin M., Longo G., Capaccioli M. et al., 2003, MNRAS, 343, 353
%
\bibitem[Schneider \& Elmegreen(1979)]{SchElm}
Schneider S. S., Elmegreen B. G., 1979, ApJ, 41, 87
%
\bibitem[Schneider, Ehlers \& Falco(1992)]{SEF}
Schneider P., Ehlers J., Falco E. E., 1992, {\it Gravitational
Lenses} (Springer Verlag)
%
\bibitem[Sciama(2000)]{Sciama}
Sciama D. W., 2000, MNRAS, 312, 33
%
\bibitem[Stodolkiewicz(1963)]{Stodol}
Stodolkiewicz J. S., 1963, Acta Astronomica, 13, 30
%
\bibitem[Tadros, Warren \& Hewett(1998)]{TWH}
Tadros H., Warren S., Hewett P., 1998, New Astronomy Reviews, 42,
115
%
\bibitem[Turner, Schneider, Burke et al.(1986)]{TSB}
Turner E. L., Schneider D. P., Burke B. F. et al., 1986, Nature,
321, 142
%
\bibitem[Vilenkin(1981)]{Vil}
Vilenkin A., 1981, PRD, 23, 852
%
\bibitem[Walker(1995)]{Walker}
Walker M. A., 1995, ApJ, 453, 37
%
\bibitem[Walker(1999)]{W99}
Walker M., 1999, MNRAS, 306, 504
%
\bibitem[Walker \& Wardle(1998)]{WW98}
Walker M., Wardle M., 1998, ApJ, 498, L125
%
\bibitem[Wardle \& Walker(1999)]{WW99}
Wardle M., Walker M., 1999, ApJ, 527, L109
%
\bibitem[Walker, Wardle \& Ohishi(2003)]{WWO}
Walker M., Wardle M., Ohishi M., 2003, ApJ, 589, 810
%
\bibitem[Zeldovich(1980)]{Zel}
Zeldovich Ya. B., 1980, MNRAS, 192, 663

\end{thebibliography}
\end{document}